\documentclass[10pt]{article}
\usepackage{graphics,epsfig}

\font\smallroman=cmr9

\font\bigroman=cmr12

\begin{document}
\title{Complementary Descriptions (PART II)\\ {\bigroman A Set of Ideas Regarding the Interpretation of Quantum
Mechanics}}

\author{Christian de Ronde}
\date{}
\maketitle \centerline {Center Leo Apostel (CLEA), and}
\centerline {Foundations of the Exact Sciences (FUND),}
\centerline {Faculty of Science,} \centerline {Brussels Free
University} \centerline {Krijgskundestraat 33, 1160 Brussels,
Belgium.} \centerline {cderonde@vub.ac.be}

\begin{abstract}
\noindent Niels Bohr introduced the concept of complementarity in
order to give a general account of quantum mechanics, however he
stressed that the idea of complementarity is related to the
general difficulty in the formation of human ideas, inherent in
the distinction between subject and object. The complementary
descriptions approach is a framework for the interpretation of
quantum mechanics, more specifically, it focuses in the
development of the idea of {\it complementarity} and the concept
of {\it potentiality} in the orthodox quantum formulation. In PART
I of this article, we analyze the ideas of Bohr and present the
principle of complementary description which takes into account
Einstein's ontological position. We argue, in PART II, that this
development allows a better understanding of some of the
paradigmatic interpretational problems in quantum mechanics, such
as the measurement problem and the quantum to classical limit. We
conclude that one should further develop complementarity in order
to elaborate a consistent worldview.\end{abstract}\bigskip
\noindent

\section*{Introduction}

In this article I would like to develop in a more technical way
the ideas presented in \cite{deRondeCDI}. More specifically, I
will turn my attention to the problem of the quantum to classical
limit through the analysis of the quantum theory of measurement
and several of the most paradigmatic phenomena in the quantum
domain.

The paper is organized in two main sections. In section 1, I
describe the different levels of description regarding the quantum
theory of measurement. Firstly, we present the `quantum
description' taking into account the distinction between classical
and quantum concepts. Secondly, we take into account the `quantum
perspective'. We continue to define the `quantum context', the
`classical context' and analyze their relation. Finally, we take
into account the `empirical' level of description. In section 2, I
study several of the interpretational problems regarding quantum
mechanics and present the answers provided by the complementary
descriptions approach. Firstly, we analyze the paradigmatic
Stern-Gerlach experiment as a more general expression of the
measurement problem and the quantum to classical limit. Secondly,
we study from this approach the Schr\"odinger cat paradox.
Thirdly, we discuss the relation between the quantum and the
classical descriptions related to the Bell inequalities and the
hidden measurement approach by the Brussels group. Finally, I
present the conclusions.

\section{Levels of Description in the Quantum Theory of Measurement}

In this section I will use the definitions presented in
(\cite{deRondeCDI}, Sect. 2) to develop the different levels of
description expressed in the quantum theory of measurement.

Planck introduced the quantum of action in 1900 producing a
complete break through the system of classical mechanics and
electrodynamics; but nevertheless he was forced to continue to use
this system. {\it ``This dual position constitutes from the
beginning an epistemological paradox. In order to overcome it or
alleviate it, it became necessary at the very least to attempt a
definite connection between the two languages which physics now
had to employ continuously and side by side."}\footnote{E.
Cassirer quoted from (\cite{Cassirer56}, p.111).} My aim in the
following section will be to expose as explicitly as possible the
independence of these two languages; showing that {\it the
classical and the quantum description are
complementary}\footnote{One could argue, following Ernst Cassirer
(\cite{Cassirer56}, p.89) that this state of affairs is quite
equivalent to that of dynamical and statistical laws: ``{\it With
statistics a new method and a new instrument of description was
introduced into science, an instrument which was proved usable and
eminently fruitful even in those areas where the existing
classical methods failed. However instead of comprehending the
significance of this new method and of appreciating its particular
logical character, instead of allowing it its due position in the
systematic totality of physical determination, exactly the
opposite path was chosen. Dynamic and statistical laws were not
regarded as two complementary methods and directions, as two
different modes of description; they were instead opposed as the
``determined" and the ``undetermined".}"}. For this purpose we
will analyze the different levels of description taken into
account by orthodox quantum mechanics. My claim is that every
level, which is determined by the concepts involved, is
independent with respect to a different level, in which the
concepts used might not be compatible.

When analyzing the measurement problem from the complementary
descriptions approach we assume there are different levels of
descriptions in which the process takes place, each of them with
the same ontological status, but with different conceptual schemes
applied. As discussed in (\cite{deRondeCDI}, Sect. 4) within
certain definite contexts the concepts can be ``used" ({\it as if}
they were the same concept) by different (complementary)
descriptions at the same time; however, when extending the meaning
of the concept to the general framework of either level
incompatibilities and paradoxes appear.

\subsection{The Quantum Description: Classical and Quantum Concepts}

I take the quantum description to be defined by its formalism,
namely, Dirac's formulation on Hilbert space \cite{Dirac} (which
is equivalent to Schr\"odinger's wave mechanics or the matrix
mechanics formulation developed by Heisenberg, Born and Jordan),
together with the interpretation.

There are two main questions regarding the interpretation of the
quantum formalism. Firstly: what is the ontology regarding the
{\it Schr\"odinger wave function}? Secondly: what is the meaning
of a {\it superposition}? These two questions are in many cases
confused and taken to be equivalent. I will try to show that each
of them has it's own independence regarding the meaning of the
quantum theory. These two questions will be further analyzed in
section 1.2 and 1.3. In the following section we will focus on the
description given by quantum mechanics and the concepts involved
stressing the incompatibilities with the concepts presented in the
classical framework.

\subsubsection{The Concepts of ``Space" and ``Time"}

The time evolution of the wave function description is completely
deterministic and expresses the possible correlations that will
appear in the selection of the different {\it possible contexts}.
Heisenberg (\cite{Heisenberg49}, p.65), Bohr
(\cite{WheelerZurek83}, pp.88-92) and Pauli (\cite{Hendry84},
p.126) made it clear that the description in terms of wave
function precludes the possibility to maintain a classical
space-time description\footnote{One should acknowledge however
that this is a question which is deeply connected with the
interpretation one is willing to make of the quantum formalism;
for example if one takes the Bohmian interpretation it is possible
to maintain a classical conception of position and trajectory.}.
This `limitation' was imposed in the description by the quantum
postulate, which states that any change in action must be an
integral multiple of Planck's constant. Thus, {\it ``[quantum
mechanic's] essence may be expressed in the so-called quantum
postulate, which attributes to any atomic process an essential
discontinuity, or rather individuality, completely foreign to the
classical theories and symbolized by Planck's constant of
action."}\footnote{Bohr quoted from \cite{Bohr34}, p.53.} It was
the acknowledgement of such limitations (as epistemological
principles) which led Heisenberg to the development of the new
quantum formulation; as Bohr explains:

{\smallroman
\begin{quotation}
``As is known, the new development was connected in a fundamental
paper by Heisenberg, were he succeeded in \emph{emancipating
himself completely from the classical concept of motion} by
replacing from the very start the ordinary kinematical and
mechanical quantities by symbols which refer directly to the
individual processes demanded by the quantum postulate." N. Bohr
(\cite{Bohr28} quoted from \cite{WheelerZurek83}, p.106, emphasis
added) \end{quotation}}

The quantum description rests on the fundamental presupposition of
a {\it discrete} description\footnote{The classical description,
on the other hand, is an expression of continuity. See for example
the article of Von Weizs\"acker: {\it Kontinuit\"at und
M\"oglichkeit} \cite{vWeizsacker51} in which he discusses the
limitations of classical and quantum descriptions.}. This does not
preclude the quantum description from using the concept of space
within a {\it definite context}; i.e. the Schr\"odinger wave
mechanics in the {\it position representation}. What remains still
not possible, is to maintain a consistent view together with the
context given by the {\it momentum representation} (as it can be
done in classical mechanics). Space and time in the quantum
description are not the same concepts as space and time in the
classical mechanics; Heisenberg (\cite{Hendry84}, p.112) was aware
of the difficulty of using such concepts in the quantum
formulation: {\it``that space and time are actually only
statistical concepts, as, perhaps, are temperature, pressure, etc.
in a gas. I mean, that space-like and time-like concepts are
meaningless for {\it one} particle, and that they become more and
more meaningful the more particles are treated."} Such difficulty
stands from the fact quantum mechanics is a theory created with
completely different epistemological principles than classical
mechanics. It was in the concept of complementarity where Bohr saw
a way to regain the space-time description and
causality\footnote{See for example \cite{Bohr28} and also
\cite{Pauli94}, Chap. 10: {\it``Space, Time and Causality in
Modern Physics"}.}.

Although there exists several possibilities in order to regain a
classical conception of space and time within the quantum
formulation\footnote{Maybe the most famous one is that presented
by David Bohm (\cite{Bohm52a} and \cite{Bohm52b}).}, I believe the
beauty of physics relies in the creation of concepts which can
open new paths in the understanding and comprehension of the
world. I will thus, not look back and go further into this
development.

\subsubsection{The Principle of Separability}

Deeply connected to the concepts of space and time stands the {\it
principle of separability}; which following Howard
(\cite{Howard89}, p.226) can be expressed:

{\smallroman
\begin{quotation}
{\bf Separability Principle}: The contents of any two regions of
space separated by a nonvanishing spatio-temporal interval
constitute separable physical systems, in the sense that (1) each
possesses its own, distinct physical state, and (2) the joint
state of the two systems is wholly determined by these separated
states.\end{quotation}}

In other words, the presence of a nonvanishing spatio-temporal
interval is a sufficient condition for the individuation of
physical systems and their associated states, then `physical
wholes' are not more than the `sums of their parts'. One of the
most important proponents of this position was Albert Einstein,
who was clearly against the holistic nature of quantum mechanics.
In a letter to Max Born dated 5 April, 1948, Einstein writes:

{\smallroman
\begin{quotation}
``If one asks what, irrespective of quantum mechanics, is
characteristic of the world of ideas of physics, one is first
stuck by the following: the concepts of physics relate to a real
outside world, that is, ideas are established relating to things
such as bodies, fields, etc., which claim a `real existence' that
is independent of the perceiving subject --ideas which, on the
other hand, have been brought into as secure a relationship as
possible with the sense-data. It is further characteristic of
these physical objects that they are thought of as arranged in a
space-time continuum. An essential aspect of this arrangement of
things in physics is that they lay claim, at a certain time, to an
existence independent of one another, provided these objects `are
situated in different parts of space'. Unless one makes this kind
of assumption about the independence of the existence (the
`being-thus') of objects which are far apart from one another in
space --which stems in the first place in everyday thinking--
physical thinking in the familiar sense would not be possible. It
is also hard to see any way of formulating and testing the laws of
physics unless one makes a clear distinction of this kind." A.
Einstein (quoted from \cite{Born71}, p.170) \end{quotation}}

However, Einstein was aware there was no logical inconsistency in
dropping this principle and accepting the holistic character of
the quantum theory. At the end of the same letter he points out
the following:

{\smallroman
\begin{quotation}
``There seems to me no doubt that those physicists who regard the
descriptive methods of quantum mechanics as definite in principle
would react to this line of thought in the following way: they
would drop the requirement for the independent existence of the
physical reality present in different parts of space; they would
be justified in pointing out that the quantum theory nowhere makes
explicit use of this requirement." A. Einstein (quoted from
\cite{Born71}, p.172)
\end{quotation}}

Einstein wrote that this was unacceptable to him because the price
to pay was to give up physical reality, which for him was placed
by the preconditions of the separability and locality principles.
As a matter of fact, maybe this was the most important difference
between the thought of Einstein and Bohr; i.e. the difference
between an epistemological and an ontological position, as Folse
argues:

{\smallroman
\begin{quotation}
``[...] Einstein realism is based upon the assumption of
separability, whereas Bohr's realism is grounded in the quantum
postulate which in effect is the very denial of Einstein's
separability assumption. A simple but suggestive way of expressing
the philosophical reason for this difference between Bohr and
Einstein is that for Einstein the ontological demands for an
objective description precede the epistemological demands, whereas
for Bohr the epistemological precede the ontological. As we have
seen, Einstein's rejection of Bohr's epistemological claim that we
have in the quantum formalism a complete description of physical
reality is based on his view that locality and separability must
be essential conditions for any concept of physical reality that
can be objectively described. Bohr does not deny that physics
describes reality, but what he does hold is that --at least if we
regard the effects of the quantum revolution as here to say--
physics cannot make these the necessary conditions for the
physical reality if that reality is to be objectively described.
Thus Bohr's concept of reality is derivative from his
epistemology, or as he would put it, from the ``epistemological
lesson" which the quantum revolution has taught us." J. Folse
(\cite{Folse87}, p.174) \end{quotation}}

Einstein's necessary conditions for making science possible
presuppose a `metaphysical science', a `classical science'; if
this would be the case every physical theory should in a late
stage reduce to classical physics. Feynman (\cite{Feynman92},
p.129) has presented this idea in the most clear way in his famous
dictum: {\it ``I think I can safely say that nobody understands
quantum mechanics."} If one presupposes that understanding must
lie within the walls of classicality then quantum mechanics will
remain un-understandable forever. But the question I would like to
pose once again is: {\it``What does one mean by `understanding'?"}

Although reductionism has provided an important guide to science
for many centuries, quantum mechanics resists such an analysis.
When a quantum state of a compound system is described by a pure
(`entangled') state, its subsystems cannot be described by a pure
state\footnote{For a study of the concept of Holism see for
example \cite{Michiel04} and references therein.}. The holistic
path opens a world of new ideas and concepts. The new quantum
technology, which embraces projects as quantum computers, quantum
teleportation and quantum cryptography, is a clear witness to this
fact.

It has been argued (\cite{Howard89}, \cite{Teller85} and
\cite{Teller89}) that all local, separable theories are
empirically false when applied to the microphysical interactions
examined in the Bell type experiments, and that if one is
unwilling, as we are, to sacrifice locality, the assumption of
separability must be recognized as the source of difficulty.
Furthermore, separable theories are fundamentally incompatible
with quantum mechanics because of their separable manner of
individuating systems and states. That quantum mechanics tells us
about the strange entangled properties of the systems it describes
is known since Schr\"odinger's 1935 paper \cite{Schrodinger35}.
But by no means this is restricted to the wave picture of the
quantum theory. Quantum logic is maybe the most general framework
in which quantum mechanics can be expressed\footnote{For a review
in quantum logic see for example the review of Dalla Chiara and
Giuntini \cite{DallaChiaraGiuntini01}.}, yet we know since Aerts
\cite{Aerts81} that even in this abstract setting the
inseparability of quantum systems is factual: {\it quantum
mechanics simply cannot describe separated systems}.

Separability is one of the cornerstones of classicality, it goes
together with the principle of non contradiction which asserts it
is wrong to think at one and the same instant that which {\it is}
and {\it is not}. We believe the path laid down by the quantum
theory goes against the idea of entity and although a bridge
(rather than a limit) must be build up between these complementary
descriptions (the classical and the quantum) the idea of entity
should not be forced into the interpretation of the quantum
formalism; rather we should focus in finding new ways in which
this character (non-separability) is exposed with all its
strength.

\subsubsection{The Concept of ``Measurement"}

The received view tells us that in classical mechanics observation
does not influence in any way the object which we study.
Measurement is in this sense ``outside" the theory; it is a way of
acquiring information without in any way perturbing the system.
This allows us to think of a reality acquirable by the senses
which is independent of the act of observation. In quantum
mechanics this is not the case, we are explicitly confronted with
the fact that what we choose to study defines our system, it
creates an entity and the elements of physical reality.

It is the epistemological grounding of these two theories which
allow us to find a totally different structure in logic and
language. In classical mechanics it is normally accepted that
whatever we choose to measure, it will not change our object,
thus, the object is completely independent of the act of
measurement. In the quantum theory the concept of measurement has
a totally different significance: the choice of the measurement we
want to perform defines the context, a context which is mutually
incompatible with other contexts; it defines the elements of
physical reality of the system and thus, it defines the {\it
entity}. There is no space in quantum physics for a `detached
observer'. Wolfgang Pauli was very much aware of this character,
as it is expressed in a letter to Niels Bohr:

{\smallroman
\begin{quotation}
``[...] it seems to me quite appropriate to call the conceptual
description of nature in classical physics, which Einstein so
emphatically wishes to retain, ``the ideal of the detached
observer". To put it drastically the observer has according to
this ideal to disappear entirely in a discrete manner as hidden
spectator, never as actor, nature being left alone in a
predetermined course of events, independent of the way in which
phenomena are observed. ``Like the moon has a definite position"
Einstein said to me last winter, ``whether or not we look at the
moon, the same must also hold for the atomic objects, as there is
no sharp distinction possible between these and macroscopic
objects. Observation cannot {\it create} an element of reality
like position, there must be something contained in the complete
description of physical reality which corresponds to the {\it
possibility} of observing a position, already before the
observation has been actually made." I hope, that I quoted
Einstein correctly; it is always difficult to quote somebody out
of memory with whom one does not agree. It is precisely this kind
of postulate which I call the ideal of the detached observer.

In quantum mechanics, on the contrary, an observation {\it hic et
nunc} changes in general the ``state" of the observed system, in a
way not contained in the mathematical formulated {\it laws}, which
only apply to the automatical  time dependence of the state of a
{\it closed} system. I think here on the passage to a new
phenomenon of observation which is taken into account by the
so-called ``reduction of the wave packets". As it is allowed to
consider the instruments of observation as a kind of prolongation
of the sense organs of the observer, I consider the impredictable
change of the state by a single observation --in spite of the
objective character of the results of every observation and
notwithstanding the statistical laws of frequencies of repeated
observation under equal conditions-- to be {\it an abandonment of
the idea of the isolation (detachment) of the observer from the
course of physical events outside himself.}

To put it in nontechnical common language one can compare the role
of the observer in quantum theory with that of a person, who by
its freely chosen experimental arrangements and recordings brings
forth a considerable ``trouble" in nature, without being able to
influence its unpredictable outcome and results which afterwards
can be objectively checked by everyone." W. Pauli
(\cite{Laurikainen88}, p.60)
\end{quotation}}

In quantum mechanics we cannot escape the fact that we are
confronted with ourselves. This is not the collapse of the wave
function, but rather, the collapse of the Cartesian cut. As
clearly expressed by Heisenberg (\cite{Heisenberg27}, p.187):
\textit{``I believe that one can formulate the emergence of the
classical `path' of a particle pregnantly as follows: \textbf{the
`path' comes into being only because we observe it.}}" The idea of
`measurement' in the classical framework, as discovering an
objective independent reality has nothing to do with the idea of
`measurement' in the quantum description which presents us not
only as discoverers but also as creators of physical
reality\footnote{This idea has been developed by Diederik Aerts in
{\it the creation discovery view}. See for example
\cite{Aerts88}.}.

\subsection{The Quantum Perspective: The Schr\"odinger Wave
Function}

In quantum mechanics the state function $|\Psi\rangle$ can be
expressed in different {\it representations}, each of which is
given in the formalism by different basis $\{B, B', B'', ...\}$.
For each representation we obtain respectively
$\{|\Psi_{B}\rangle, |\Psi_{B'}\rangle,
|\Psi_{B''}\rangle,...\}$\footnote{More generally one can think in
terms of density operators: firstly a $\rho$ without a definite
basis, and secondly, $\{\rho_{B}, \rho_{B'}, \rho_{B''},...\}$
given by the density operator in each basis $\{B, B', B'',
...\}$.}. {\it We have to choose in which basis we are going to
write the wave function.} Each {\it representation/basis}
expresses a {\it context} which can be in principle incompatible
to a different context. This is were all the trouble starts:
compatibility\footnote{For an analysis of the concept of
compatibility see for example the very interesting passage of the
book of Asher Peres \cite{Peres93}, Chap. 7.}. It has been proved
by Kochen and Specker \cite{KochenSpecker67} that in a Hilbert
space d$\geq$3, it is impossible to associate numerical values, 1
or 0, with every projection operator $P_{m}$, in such a way that,
if a set of {it commuting} $P_{m}$ satisfies $\sum P_{m}=\amalg$,
the corresponding values, namely ${\it v}(P_{m})=0$ or 1, also
satisfy $\sum {\it v}(P_{m})=1$. This means that if we have three
operators A, B and C, where $[A,B] = 0$, $[A,C] = 0$ but $[B,C]
\neq 0$ it is not the same to measure $A$ alone, or $A$ together
with $B$, or together with $C$\footnote{For an analysis of the
Kochen-Specker theorem see for example \cite{Bub97} and
\cite{Bacciagaluppi96}.}. It is the incompatibility of contexts;
i.e. the impossibility of assigning a global truth valuation to
the projection operators of different contexts, which brings into
stage the concept of {\it choice}. In classical mechanics, due to
the compatible structure, one can neglect this level. Classically
the choice of the context {\it discovers} (rather than {\it
creates}) an element of physical reality, which of course was
already there... just as the moon is outside there without us
having to chose anything at all.

Although one can still take an interpretation in which there is a
preferred basis\footnote{Take for example the Kochen-Dieks modal
interpretation which relies on the Schmidt decomposition.} the
choosing is then transferred to the {\it factorization} which
defines the system. We cannot escape the fact that {\it we always
have to make a choice in order to define our system; the state
function $\rho$ is the condition of possibility for this choice to
take place.} From our view point the indefinite (without a
definite representation/basis) wave function expresses a {\it
perspective}, the precondition that makes possible the {\it
choice}. As David Finkelstein has argued:

{\smallroman
\begin{quotation}
``To speak of `the wave function of the system' is a syntactic
error. A wave function is not a property of the system in any
classical sense, but gives far more information about the
experimenter. [...] A wave function tells us more about the {\it
act} of measurement than about the result. It is a verb, not a
noun; to treat it as a thing is a mistake in syntax. Having
committed this error, one is forced to follow with another, the
idea of the collapse of the wave function. It is not `the' wave
function that `collapses.' It is, rather, that first we do one
thing, and then another." D. Finkelstein (\cite{Finkelstein87},
p.292)
\end{quotation}}

It is interesting to take into account the problem of objectivity
within this approach. Quite often `choice' is regarded as a
subjective element of which the exact sciences might be better
with out. As D'Espagnat points out:

{\smallroman
\begin{quotation}
``Of course the statements that compose physics are objective. All
are not so in the same way however. Many, especially in classical
physics, have such a form that, at least at the times of classical
physics, they could be understood by a conventional realist as
faithfully describing attributes (or existence) of ``objects as
they really are". Let them be called ``strongly objective". Others
explicitly or implicitly involve in their very wording some
reference to human actions, abilities and, last but not least,
perceptions. Let them be called ``weakly objective". Weakly
objective statements, especially in the form of predictive
observational rules, play quite a specially important role in
quantum mechanics." D'Espagnat (\cite{D'Espagnat98}, p.7)
\end{quotation}}

Following Bernard D'Espagnat's definitions the {\it perspective}
can be thus considered as {\it weakly objective}. It presents the
possibility of choosing between different re. Once the basis has
been chosen a definite {\it context} enters the scene and the
quantum entity is created. Furthermore, {\it strong objectivity}
is regained in the form of statistical causality\footnote{Wolfgang
Pauli expressed that the idea used in science of {\it ``nothing
happens without a cause"} no longer applies to individual events
in quantum mechanics and that only mean values have rationally
analyzable causes. One is able to recover the idea of causality in
terms of a `statistical causality' by taking into account a series
of measurements.}. It is `choice', which appears explicitly in
quantum mechanics, what makes us face the problem of potentiality.

\subsubsection{The Concept of ``Ontological Potentiality"}

The idea of regarding actuality as the real is a heavy burden in
western though. Contrary to this position, I would like to
introduce the concept of {\it ontological
potentiality}\footnote{This concept is presented more extensively
in \cite{deRonde05a}.} which presents us with the most important
character of potentiality, that of describing, as Aristotle
expressed, a different form of the being. Aristotle distinguishes
between two types of potentiality. Firstly, {\it generic
potentiality}: the potentiality of a seed that can transform into
a tree; this generical sense is not that which interests Aristotle
but rather the second possibility, the {\it potentiality as a mode
of existence}: the poet has the capacity of not writing poems. It
is not only the potentiality of doing this or that thing but
rather the potentiality of not-doing, potentiality of not being,
of not passing to the actual. What is potential is capable of
being and not being. This is the problem of potentiality: {\it the
problem of possessing a faculty.} What do I mean when I say ``I
can", ``I cannot".

{\smallroman
\begin{quotation}
``What is essential is that potentiality is not simply non-Being,
simple privation, but rather the existence of non Being, the
presence of an absence; this is what we call a ``faculty" or
``power." ``To have a faculty" means to have a privation. And
potentiality is not a logical hypostasis but the mode of existence
of this privation. But how can an absence be present, how can a
sensation exist as anesthesia? This is the problem that interests
Aristotle." G. Agamben (\cite{Agamben99}, p.183)\end{quotation}}

In quantum mechanics we are faced with the concept of choice. The
quantum state expresses the potentiality of choosing a definite
experimental arrangement, it is the condition of possibility for
an action to take place. This finds its limit in the mode of
existence of a privation, a being which {\it is} and {\it is not}
at the same time. The central point of the concept of {\it
ontological potentiality} is that it cannot be reduced to {\it
actuality} and presents us with a different form of the being,
i.e. {\it the necessity of considering potentiality as
ontologically independent of actuality}. I think that in order to
reach a deep understanding of the quantum theory it is necessary
to go further and develop this concept which allows us to think
quantum mechanics in a new fresh way. Wolfgang Pauli had foreseen
this path and pointed directly to it:

{\smallroman
\begin{quotation}
``Science today has now, I believe, arrived at a stage were it can
proceed (albeit in a way as yet not at all clear) along the path
laid down by Aristotle. The complementarity characteristics of the
electron (and the atom) (wave and particle) are in fact
``potential being," but one of them is always ``actual nonbeing."
{\it That is why one can say that science, being no longer
classical, is for the first time a genuine theory of becoming and
no longer Platonic.}" W. Pauli (\cite{Pauli94}, p.46, emphasis
added)\end{quotation}}

\subsubsection{Entity in the Quantum Perspective}

In the quantum perspective we have the {\it undetermined wave
function} $|\Psi\rangle$; i.e. the wave function without a
determined representation. Before the choice has taken place it is
not possible to define the entity. The choice defines explicitly
the quantum entity. It is wrong to think the wave function
expresses a subject (as opposed to an object) but rather {\it it
expresses the condition of possibility for the act of
subjectifying}. There is no entity in the quantum perspective,
nothing like a wave function of the universe in any mechanistical
sense is possible. Those who speak of the wave function of the
universe seek for an objective interpretation of the quantum
formalism, however, if such thing as the wave function of the
unverse would exist, {\it compatibility} and {\it choice} would be
superfluous turning quantum mechanics into a classical scheme
\cite{AertsdeRondeBart}.

\subsection{The Quantum Context: The Superposition State}

Once we choose a representation; i.e. a basis or factorization, we
define the quantum entity, which is expressed by the Schr\"odinger
wave function in a definite basis $|\Psi_{B}\rangle$ or more
generally by a density operator $\rho_{B}$. In this level there is
potentiality in an ontological sense, although there is a
distinction with that one explained above, there is no choice
involved in this level and, in this sense, {\it strong
objectivity} is regained in the form of statistical causality.

If quantum mechanics would be talking only about ensembles of
systems this would be the end of the story. Contrary to this
position, we take quantum mechanics to describe individual
entities represented by density operators. Furthermore we stay
close to the idea that the probability expressed in the formalism
has nothing to do with ``lack of knowledge".

{\smallroman
\begin{quotation}
``It was wave or quantum mechanics that was first able to assert
the existence of {\it primary probabilities} in the laws of
nature, which accordingly do not admit of reduction to
deterministic natural laws by auxiliary hypotheses, as do for
example the thermodynamic probabilities of classical physics. This
revolutionary consequence is regarded as irrevocable by the great
majority of modern theoretical physicists --primarily by {\it M.
Born, W. Heisenberg and N. Bohr}, with whom I also associate
myself." W. Pauli (\cite{Pauli94}, p.46)\end{quotation}}

\subsubsection{Potentiality}

Heisenberg was very close to Platonic ideas\footnote{For an
exposition of his though see for example \cite{Heisenberg58},
\cite{Heisenberg72} and \cite{vWeizsacker85}.}. From his young
readings on the Timaeus one can trace his idea that we must think
of the micro-world as constituting of two different levels. The
first level of {\it potentialities} which is independent of
observation and the second constituted by {\it actualities} which
we attain in our observation. The quantum mechanical formalism,
and particularly its wave function, describes the level of
potentialities.

{\smallroman
\begin{quotation}
``In the experiments about atomic events we have to do with things
and facts, with phenomena that are just as real as any phenomena
in daily life. {\it But atoms or the elementary particles are not
as real}; they form a world of potentialities or possibilities
rather than one of things or facts" W. Heisenberg
(\cite{Heisenberg58}, p.160, emphasis added).\end{quotation}}

Heisenberg still regarded the superposition from a positivistic
view in which potentiality is thought as possible actualities.
Where reality is again related to the actuality of an experimental
observation. Contrary to this position, within the complementary
descriptions approach atoms and elementary particles, as described
by quantum mechanics; i.e. by superpositions, {\it are just as
real} as what we are used to call things or facts. Furthermore the
potentiality expressed in the quantum superposition is
ontological. A superposition expresses a unity, a single system in
a different level to that of actuality. Thus, in the quantum
context potentiality is not expressed as a faculty, which is the
expression of potentiality in the perspectival level, but rather
as an expression of the potential in a dialectical sense. Thus,
being and non-being should not be read in a propositional form,
but rather, in a dialectical manner of which becoming is the
synthesis.

\subsubsection{Entity in the Quantum Context}

In the quantum context we recover the concept of entity: a pure
state in a definite representation, $|\Psi_{B}\rangle$, or more
generally, an {\it improper mixture}\footnote{See
\cite{D'Espagnat76}, Chap. 6 for an explanation of the distinction
between {\it proper} and {\it improper} mixture.} given by a
density operator $\rho_{B}$. An {\it improper mixture} denies the
possibility of applying an ignorance interpretation to its
elements: it is not possible to think the system {\it is} in one
of the states, but due to our lack of knowledge, there is
ignorance with respect to which. This impossibility makes us face
the holistic character of quantum mechanics and the need of an
interpretation which takes into account the unity of a
superposition. I am willing to go still further and regard this
quantum entity as ontologically existent. A superposition
expresses a different form of the being in terms of ontological
potentiality, Schr\"odinger cats being a clear example of this
character. We will go in more detail in this analysis in Sect.
2.2.

\subsection{The Classical Context: Statistical Mechanics and Elements of Physical Reality}

Pauli repeatedly emphasized, that once the measuring apparatus is
chosen, objective reality holds in the sense that the observer
ceases to have an influence on the outcome of the experiment and
reproducibility of the result is obtained in the form of objective
probability laws for series of repeated measurements:

{\smallroman
\begin{quotation}
``...that in the systematic foundation of quantum mechanics one
should start, more than it is done so far (e.g. Dirac), from the
composition and separation of systems. For this is --as Einstein
rightly felt-- a very fundamental point of quantum mechanics which
in addition, has a direct connection with your considerations
about the cut and the possibility of its displacement in the spot"
W. Pauli (quoted from \cite{Enz85}, p.245)
\end{quotation}}

The point stressed by Pauli pins the interpretational step which
originates several of the paradoxes emergent in quantum mechanics;
i.e. the relation between the {\it quantum context} and the {\it
classical context}. To reconcile these complementary views one
needs an interpretational jump which can bring together the one
and the many. The solution to this paradox has been developed by
Bas Van Fraassen who was able to recover the seemingly paradoxical
relation between a {\it proper} and an {\it improper mixture}
through the distinction of a {\it dynamical state} and a {\it
value state}. With this interpretational development it is
possible to recover an objective probability in terms of ignorance
without any inconsistency, this is from our view the {\it
interpretational jump}\footnote{I use the idea of {\it jump} as
opposed to that of {\it limit}.} needed in order to go from the
{\it quantum context} to the {\it classical context}.

{\smallroman
\begin{quotation}
``...the interpretation given to density operators is in terms of
ignorance: [...] the interpretation gives the probabilities of
various possible states of affairs of which it is assumed that
only one is actually realized. From the point of view of the
standard interpretation this combination of improper mixture and
an ignorance interpretation may seem paradoxical. To resolve the
paradox, it is crucial to notice the distinction which is made in
the modal interpretation between physical properties and
mathematical states. The latter are defined in Hilbert space and
encode probabilistic information about the properties (values of
physical magnitudes) which are present in the system itself. There
is no one-to-one relation between physical properties and
mathematical states. Now, the ignorance inherent in the modal
interpretation pertains to physical properties; not to
mathematical states. It is not assumed that the description by
$\rho$ is incomplete and that some other state (namely a pure
state) would give a more complete description. The modal
interpretation is therefore only in a quite specific sense similar
to the standard ignorance interpretation and the density operators
are partial traces, and in this sense improper mixtures. They
cannot be considered as incomplete mathematical state
specifications (in contradiction to proper mixtures in the
standard approach). But there still is room for ignorance about
the values of physical magnitudes!" D. Dieks and P. Veramaas
(\cite{DieksVermaas95}, p.155)
\end{quotation}}

From the principle of complementary descriptions this distinction
is taken to its last consequences and determines two
ontologically, complementary states; the {\it holistic state} and
the {\it reductionistic state}\footnote{In the modal
interpretation the ``holistic state" (superposition state)
reflects the {\it potential mode of the being}; while the
``reductionistic state" (proper classical mixture) reflects the
{\it elements of physical reality}. Van Fraassen calls them {\it
dynamical} and {\it value} state, while Dieks calls them {\it
mathematical} state and {\it physical} state, respectively. The
distinction in terminology is not superfluous but expresses the
philosophical attitude towards the interpretation of the theory.
On the one hand, Van Fraassen shows his empiricist attitude by
assigning ``reality" to the value of the actual term. Dieks, on
the other hand, makes a clear distinction between the levels of
discourse: mathematical {\it vs.} physical. I would like to
emphasize the distinction between {\it concepts} and {\it things
as they are}; in this sense holistic and reductionistic states are
nothing more and nothing less than {\it useful descriptions}.}
 which are ontologically equivalent but belong to the quantum and
the classical levels respectively\footnote{It is interesting to
notice that the holistic and reductionistic state can be regarded
as particular expressions of the idea of potentiality as expressed
by Aristotle; i.e. potentiality as a faculty (holistic state) and
potentiality as becoming (reductionistic state).}. At this point I
escape the orthodox `empiricist view':

{\smallroman
\begin{quotation}
``To be an empiricist is to withhold belief in anything that goes
beyond the actual, observable phenomena, and to recognize no
objective modality in nature. To develop an empiricist account of
science is to depict it as involving a search for truth only about
the empirical world, about what is actual and observable. Since
scientific activity is an enormously rich and complex cultural
phenomenon, this account of science must be accompanied by
auxiliary theories about scientific explanation, conceptual
commitment, modal language, and much else. But it must involve
throughout a resolute rejection of the demand for an explanation
of the regularities in the observable course of nature, by means
of truths concerning a reality beyond what is actual and
observable, as a demand which plays no role in the scientific
enterprise." B. van Fraassen (\cite{VanFraassen80},
pp.202-203)\end{quotation}}

The idea of Van Fraassen is that: {\it There is only one actual
reality. Modalities are in our theories, not in the
world}\footnote{See also \cite{VanFraassen80}, Sect. 5.}. This is
close to the position supported by Bohr, whom regarded the
Schr\"odinger wave function as an {\it
abstraction}\footnote{{\it``The symbolical character of
Schr\"odinger's method appears not only from the circumstance that
its simplicity, similarly to that of the matrix theory, depends
essentially upon the use of imaginary arithmetic quantities."}
(Bohr, quoted from \cite{WheelerZurek83}, p.111).} without any
ontological meaning. Contrary to this view within the
complementary descriptions framework we regard the quantum
description from an equal ontological standpoint to the classical
one. In this sense the formalism of quantum mechanics, and
therefore a superposition state reflects a character of reality
just in the same way a Stern-Gerlach apparatus does; there is no
{\it actuality} voided of description.

It is important to notice that modal interpretations, as they
stand, do not talk about `Schr\"odinger cats', {\it
superpositions}, as ontologically existent entities\footnote{This
is being empirically tested an all the experiments seem to point
out that Schr\"odinger cats, as individual entities, do exist! See
for example \cite{Friedman02}.}, they only talk about {\it actual
cats!} I wish to stand for the opposite: {\it Modalities are in
the world, not only in our theories}, they express a complementary
description of reality\footnote{It is interesting to point out
that van Fraassen's position is very close to  that of Bohr, who
pointed out the importance of the classical concepts, which
defined the phenomena, and he repeatedly stressed the fact that
the wave function should only be considered as an abstraction.}.
This determines an empirical distinction between my interpretation
of quantum mechanics and those which regard superpositions as void
of an ontological status. The consequence of such position with
respect to modal interpretations will be further analyzed in
future articles (\cite{deRonde05b} and \cite{deRonde05c}).

\subsection{The Empirical Description: Observation and Reference}

In his famous `cat article' of 1935 Schr\"odinger writes the
following:

{\smallroman
\begin{quotation}
``Reality resists imitation trough a model. \textbf{So one lets go
of naive realism and leans directly on the indubitable proposition
that \textit{actually} (for the physicist) after all is said and
done there is only observation, measurement.} Then all our
physical thinking thenceforth has the sole basis and as sole
object the results of measurements which can in principle be
carried out, for we must now explicitly {\it not} relate our
thinking any longer to any kind of reality or to a model. All
numbers arising in our physical calculations must be interpreted
as measurement results. But since we didn't just now come into the
world and start to build up our science from scratch, but rather
have in use a quite definite scheme of calculation, from which in
view of the great progress in Q.M. we would less than ever want to
be parted, we see ourselves forced to dictate from writing-table
which measurements are in principle possible, that is, must be
possible in order to support adequately our reckoning system."
 Schr\"odinger (\cite{Schrodinger35}, p.156, emphasis added) \end{quotation}}

It is important to notice that, even though one is not willing to
remain a {\it naive realist}, one should as well reflect on the
idea that observation and measurement are (even for the
physicists) {\it indubitable propositions}. I am not wiling to
give this metaphysical presupposition for granted. This is the
problem of the justification of experimental observation which is
as well the problem of subjectivity. If one does not rely on
ontology, observation looses content, meaning. There is no
possible observation without an ontology on which the phenomena
can be embedded\footnote{Michel Bitbol \cite{Bitbol00} has also
called the attention upon the relation between description and
theory through the name epistemological circle: {\it ``Par `cercle
epistemologuique' j'entends la relation r\'eciproque
auto-consistente entre une theorie scientifique et la maniere dont
celle th\'eorie conduit a se figurer en retour le processus par
lequel il a \'et\'e possible de la formuler et de l'attester."}}.
Einstein pointed out this to Heisenberg {\it ``it is only the
theory which decides what can be observed."} This remark, as
Heisenberg recalls in his own autobiography, {\it Der Teil und das
Ganze}, was the idea that lead him to the uncertainty relations.

Before considering the question of Einstein: What can be
considered as an element of physical reality in the theory? We
must go back to the problem posed to some extent by Bohr: What is
given as empirically adequate? What is a phenomenon? What are the
conditions by which I can access a phenomenon?

In the kantian approach experimental observation presupposes
certain preconditions, certain categories and forms of intuition
which allows us to grasp a certain phenomena from the manifold of
data. In quantum mechanics there is an indeterministic jump from
the possible to the actual by which we are able to recover our
classical space-time description, and thus, a separated entity
(the spot on the photographic plate). However, the character of
this `problematic jump' should be clearly specified; this is to
great extent the question of which are the preconceptions stated
in the phrase: {\it a black spot appears in the photographic plate
of a Stern-Gerlach apparatus at the end of the experiment} (this
will be analyzed more carefully in section 2.1).

The complementary descriptions approach does not evade any
ontological commitment. It is in this sense a position half way
between Einstein's ontic view and Bohr's epistemological attitude
--we are able to maintain a plurality of ontologies and at the
same time take into account the epistemological problems which are
determined by such an election. Within the complementary
descriptions approach we consider the Stern-Gerlach apparatus
(with all its meaning and symbolism) as a description in itself
not more, not less objective than the one presented in the quantum
mechanical formalism in the Hilbert space. Dieks is correct to
conclude that if one is willing to assume certain realism upon the
description one develops, the structure of this description must
be taken seriously:

{\smallroman
\begin{quotation} ``The quantum realism [...] provides
us with an abstract structural description. The irreducibly
stochastic level of observations is subsumed under a theoretical
framework with a deterministic evolution law; its structure is the
non-classical structure of a Hilbert space. The world has, if we
accept elementary quantum mechanics as descriptive, this same
structure." D. Dieks (\cite{Dieks89}, p.1418)
\end{quotation}}

In relation to this distinction, one must also acknowledge that
the {\it Stern-Gerlach in my laboratory} and the {\it name} which
we use to make {\it reference} to it: ``Stern-Gerlach" are not the
same. It is the limitation in language which must be acknowledged
when discussing the interpretation of a theory. In the quantum
description as well as in the classical description {\it ``the
limits of my language mean the limits of my
world"}\footnote{Wittgenstein quoted from \cite{Wittgenstein},
Sect. 5.6.}:

{\smallroman
\begin{quotation}
``Newtonian mechanics, for example, imposes a unified form on the
description of the world. Let us imagine a white surface with
irregular black spots on it. We then say that whatever kind of
picture these make, I can always approximate as closely as I wish
to the description of it by covering the surface with a
sufficiently fine square mesh, and then saying of every square if
it is white black or white. In this way I shall have imposed a
unified form on the description of the surface. The form is
optional, since I could have achieved the same result by using a
net with triangular or hexagonal mesh. Possibly the use of a
triangular mesh would have made the description simpler: that is
to say, it might be that we could describe the surface more
accurately with a coarse triangular mesh than with a fine square
mesh (or conversely), and so on. The different nets correspond to
different systems for describing the world. Mechanics determines
one form of description of the world by saying that all
propositions used in the description of the world must be obtained
in a given way from a given set of propositions--the axionms of
mechanics." L. Wittgenstein (\cite{Wittgenstein}, pp.81-82, Sect.
6.341)\end{quotation}}

We have presented a scheme were there is a complementary relation
between the {\it actual} observation that takes place and the {\it
statistical} description with which one is able to immerse this
particular result in the {\it classical context}. As it is well
known Bohr expressed repeatedly the necessity of always using
classical concepts in order to describe experimental
setups\footnote{Bohr (quoted from \cite{WheelerZurek83}, p.4)
maintained that: {\it ``The experimental conditions can be varied
in many ways, but the point is that in each case we must be able
to communicate to others what we have done and what we have
learned, and therefore} the functioning of instruments must be
described within the framework of classical physical ideas."
(emphasis added).}. This {\it necessary condition} can be
understood within the present scheme as the necessary condition
for the quantum context to allow a meeting point through the
classical context with the actuality of an experimental outcome.

To summarize, I have presented different levels of description
which take place in the process of the quantum measurement, the
{\it quantum perspective} described by the indefinite
Schr\"odinger wave function, the {\it quantum context} described
by the quantum superposition (an improper mixture); the {\it
classical context} described by a proper mixture in which
superpositions are neglected and we regain statistical causality
as well as the elements of physical reality; and finally, the
empirical world, that which pertains to our personal experience.
In the following section I will try to show more clearly, through
some of the most paradigmatic experiments of quantum theory, how
these levels are related.

\section{Complementary Descriptions and Quantum Mechanics}

I would like now to approach from the principle of complementary
descriptions the more general problems regarding the
interpretation of quantum mechanics. Firstly, I will analyze the
Stern-Gerlach experiment. Secondly, I will address the
Schr\"odinger cat paradox. Last but not least, I will analyze the
Bell inequalities and the classical to quantum limit in the light
of the hidden measurement approach developed by the Brussels
group.

\subsection{The Neverending Stern-Gerlach Experiment}

Let's discuss in some detail the Stern-Gerlach experiment in the
light of the definitions expressed above. If we choose to describe
the apparatus classically, we must consider an object with a
definite space time position; in this case we describe the magnet
and the photographic plate with {\it classical concepts}. One must
agree that the Stern-Gerlach apparatus, as Bohr repeatedly
emphasized, is nothing but a classical object with a definite
position in space and time, with all the monadic properties which
make this object a Stern-Gerlach apparatus. What I want to stress
is that already talking about a Stern-Gerlach apparatus
presupposes a description, namely, a {\it classical} one. However,
we could also choose to describe our experiment in a ``purely"
quantum mechanical way. If we take for instance a closed room with
the Stern-Gerlach apparatus inside we must also take into account
the great number of particles composing the magnet, the
photographic plate and even the photons and particles of air in
the room\footnote{One should be aware of the phrase of Bohr {\it
``we are suspended in language."} We have to speak using our
language, which has not yet been developed to take into account
quantum mechanics, in quantum mechanics it is meaningless to speak
of ``particles" in the classical sense, however, in this case we
are forced to push the words to their limits.}. Clearly this
description exceeds by far the skills of the most capable and
obsessive experimental physicist, anyhow it should be possible in
principle as it was once proposed by Laplace in classical
mechanics. What should be clearly noticed is that this, and only
this would be the pure {\it quantum mechanical description},
namely, ``the description of the whole room by a quantum density
operator $\rho$". However, one must acknowledge that this phrase
is already a paradox in itself, a mixture of two incompatible
languages. Talking about a room with an apparatus and so on,
again, presupposes a classical description; classical objects
exist in themselves, independently of any choice. This is not true
for quantum mechanics.

Now, what is normally done when analyzing this experiment, without
any explicit remark, is an implicit coarse graining within the
{\it quantum description}: one makes a cut between an {\it
elementary quantum state} and a {\it classical pointer state}. In
this subtle point that we take often for granted in the study of
measurement interaction is the malefic seed which will grow up to
a quantum jump. If we would describe the experiment truly quantum
mechanically, we must be aware that the quantum mechanical state
which describes the complete room in which we have the system, the
apparatus, and the rest of the particles (called commonly
environment) will always {\it evolve unitary in time} with the
Schr\"odinger equation of motion:

\begin{equation}i\hbar\frac{\partial}{\partial t}\rho=[H,\rho]
\end{equation}

This means that although some correlation\footnote{It is important
to notice that quantum mechanics only provides correlations and
nothing like {\it definite states}; whatever that might mean.} may
lead, at some definite time, to correlations which might be
related to a black spot in one of the sides of the photographic
plate, at a future time, due to the infinite evolution of the
quantum mechanical state, this black spot correlations might
`reconstruct' themselves from the particles and photons in the
air, and could `whiten' back into the initial state of the
photographic plate. Even worse, the black spot, if the
experimenter is sufficiently patient (maybe $10^{50}$ years?),
might appear in the lower side! But why should we then consider
the experiment is finished, if we have shown, that if we describe
things quantum mechanically, there is no such end in the act of
measurement (as the state keeps evolving)? {\it The only things we
have are potential quantum mechanical correlations evolving till
the end of time.}

The point is that the quantum mechanical description is not very
practical when taken to its lasts fundamental character, what we
normally do is to take some crucial decisions for all practical
proposes (FAPP). First, what are the degrees of freedom of
interest, and second, when does the experiment end. The first
implies a choice in the representation, a path from the {\it
quantum perspective} to the {\it definite context}. This choice
makes possible to relate the quantum state (in the quantum
context) with the pointer state of the apparatus (in the classical
context); this cut defines which is the system and which is the
apparatus\footnote{It substantializes our quantum description by
defining entities (elements of physical reality). Substance in
place of correlations; stability instead of movement.}. It is
important to notice that already here we are dealing with an {\it
interpretational jump} relating both quantum and classical
descriptions. In the second step, we choose the instant in which
the experiment is finished; this is not a `real fact of nature'
but rather a `psychological illusion' brought into by the
structure with which we comprehend the classical world (and the
fact that we cannot wait so long, maybe $10^{50}$ years, to see
the miraculous change of the black spot from the upper side to the
lower side). These two choices hide the quantum jump and the
relation between the quantum and classical descriptions.

Ren\'e Descartes, in his Regulae, rules for the direction of mind,
presented an identification between {\it `truth'} and {\it`clear
and distinctive idea'}. Modern science has taken this distinction
to its most impressive results leading to a definite cut between
object and subject. However, one should not disregard that the
problem of objectivity, this is, the impossibility of giving a
justification of such a relation between an {\it empirical
experience} and {\it truth} remains as powerful and vivid as
before. To this point science has presented a dicovery-view of
objective truths, neglecting the creative element which is at
stage in every act of observation. As Bohr repeatedly stressed,
this is the epistemological lesson that quantum mechanics taught
us, that: {\it ``We are not merely observers, but also actors, in
the great drama of the world."}

\subsection{Schr\"odinger cats do exist!}

I would like now to consider the very interesting approach by
Dennis Dieks \cite{Dieks88a} with respect to the `Schr\"odinger's
cat' paradox \cite{Schrodinger35} and go still further in the
conclusion which should be addressed by the complementary
descriptions approach.

Quantum mechanics predicts the following state for the combined
system of atom and cat:

\begin{equation}
\left| Cat\right\rangle =\frac{1}{\sqrt{2}}(\left| up\right\rangle
\left| Dead\right\rangle +\left| down\right\rangle \left|
Alive\right\rangle) \end{equation}

The peculiarity of quantum mechanics shows itself in the further
conclusion that the combined system of closed box and its contents
is represented by a coherent superposition. The fact that the cat
is dead or alive is taken to imply, in the orthodox
interpretation, that only $(\left| up\right\rangle \left|
Dead\right\rangle)$ or  $(\left| down\right\rangle \left|
Alive\right\rangle)$ can be the correct theoretical descriptions.
It is assumed by Schr\"odinger that what we call physical systems
in everyday language must be represented in the quantum mechanical
formalism by individual state vectors, in close analogy to the
familiar practice of classical mechanics where ``physical systems"
are represented by points in a phase space. Dieks points out in
this respect that:

{\smallroman
\begin{quotation}
``The language that we use in our observation reports is the
everyday language, refined by classical physics. In this sense we
use concepts like ``electrical current", ``charge", and
``particle". Of course, such terms carry the connotations
associated with them by classical physics. But it has to be noted
that in a restricted experimental context not all the components
of the meaning given to these concepts by classical physics are
relevant and actually accessible to observation [...] the position
of a ``particle" is actually measured whereas momentum remains
``hidden" in the experiment in question..." D. Dieks
(\cite{Dieks88b}, p.1414)
\end{quotation}}

Furthermore, what should be noticed is that a language determines
in itself an ontology, and that quantum mechanics, as a
mathematical language will not be necessarily connected with the
{\it classical language}; on the contrary we have been pointing
out that what is lacking is an adequate semantical language which
can express correctly the concepts hidden in the quantum
formalism. There is a stubbornness in modern physics by which once
and again we try to explain quantum mechanics with the `old'
classical concepts. This would be tantamount to explain general
relativity only with the Newtonian concepts of absolute space and
time, the reader might already feel the bunch of beautiful
paradoxes that would arise.

The empirical observation that the cat is dead or alive does not
automatically entail a particular theoretical representation.
According to the interpretational rule given by the modal
interpretation, the observation that the cat is dead or alive,
both support the theoretical description by means of the
superposition:

{\smallroman
\begin{quotation}
``It is the state vector which is a superposition, not the cat
itself. ``State vector" and ``cat" are two concepts at different
levels of discourse..." D. Dieks (\cite{Dieks88a}, p.189)
\end{quotation}}

We can see already that this approach fits perfectly with the
complementary descriptions approach. On the one hand the
superposition is a mathematical formulation which lies in the {\it
quantum context}. On the other hand, a cat belongs to a {\it
classical account of reality}. The distance I take from most of
the modal theorist (for example \cite{DieksVermaas98},
\cite{VanFraassen91} and \cite{Vermaas99}) is an ontological one.
While most of them consider the Schr\"odinger cat to be a
theoretical entity I consider the quantum superpositions to be
ontologically independent. The {\it superposition} described by
quantum mechanics should be regarded as an entity not less
objective than the concept `cat' which is used to make {\it
reference} to the cat which is sitting right in front of my desk.
Once again one should clearly acknowledge from which level one is
speaking; it is important to stress in this sense that  the {\it
name} ``cat" is a concept, nothing more nothing less, it does not
capture {\it the thing}. The concept ``cat" (which is certainly
not the cat in my desk) and the concept ``superposition" (which is
certainly not the cat in my desk) pertain to {\it complementary
languages}. It is through languages and descriptions with which we
can make reference and create reality.

I argue that an interesting interpretation of the quantum
superposition relies on the concept of potentiality; assuming, as
we do, an {\it ontological potentiality} we are able to sustain
the interpretation of such a non-classical entity without the
common reduction into actuality. A superposition {\it is} and {\it
is not} at the same time, it denies the principle of non
contradiction opening a new descriptive level which should be
further analyzed and developed\footnote{See for example
\cite{DallaChiaraGiuntini01} were the it is clearly stated the
importance of paraconsistent logics within the formulation of
quantum mechanics.}. The {\it quantum superposition} and the {\it
classical cat} are two entities in two different levels of
description, these are not necessarily in contradiction, but only
together can give us a better understanding of the relations
between all the possible conceptual schemes that can be used to
describe reality.

\subsection{Aerts' Non-Kolmogorovian Macroscopic System}

It is well known that in the category of probability models
Accardi has given a definition of a {\it Kolmogorovian probability
model}, which is the probability model of a classical system, and
a {\it quantum probability model} which is the probability model
of a quantum system. These two probability models have a
completely different structure. It was believed for long that such
a structural relation was the main characteristic between the
classical macroscopic world and the quantum microscopical world.

Accardi was able to derive a set of inequalities (which can then
be compared with experimental results) that characterize the
Kolmogorovian model and shows that these inequalities can always
be violated by experiments with a quantum system. At the same time
Diederik Aerts \cite{Aerts81} was able to build a macroscopic
classical system that violates Bell inequalities. Since Accardi
had shown that Bell inequalities are equivalent to inequalities
characterizing a Kolmogorovian probability model and Aerts had
proved that it was possible to construct a classical system which
violated the Bell inequalities, Diederik Aerts had given an
example of a ``classical" system having a non-Kolmogorovian
probability model. Aerts (\cite{Aerts86}, p.203) points out: {\it
``This was very amazing, and the classification made by a lot of
physicists of a microworld described by quantum mechanics and a
macroworld described by classical physics was challenged
completely"}. Aerts showed that:

{\smallroman
\begin{quotation}
``If we have a physical system S and we have a lack of knowledge
about the state of S, then a theory describing this situation is
necessarily a classical statistical theory having a classical
Kolmogorovian probability model. If we have a physical system S,
and a measurement {\it e} on this physical system S, and the
situation is such that we do not have lack of knowledge about the
system S, but we do have lack of knowledge about the measurement
{\it e}, then we cannot describe this situation by a classical
statistical theory, because the probability model that arises is
non-Kolmogorovian." D. Aerts (\cite{Aerts86}, p.203)
\end{quotation}}

When I got to the Brussels group and learned about all these
results I was completely shocked. To some extent I was confronted
with my earlier thought about the concept of complementarity. I
would like to read all these results now from the principle of
complementary descriptions: the work of Aerts shows that the
entity under consideration will be a quantum or classical object
regardless of its macroscopic or microscopic `shape' but from the
description one chooses to make of the entity. From this, we can
conclude that macroscopicity and microscopicity are not the main
characteristics that determine the quantum and the classical but
rather that quantum and classical are two different and
complementary levels of description.

{\smallroman
\begin{quotation}
``If we accept our explanation for the probabilities of quantum
mechanics, namely that they are due to a lack of knowledge about
measurements, then these probabilities are not more ontological
than ordinary probabilities. They form a non-classical probability
model because they correspond to a different physical situation,
namely the physical situation where we lack knowledge about
measurements and not about the state of the system. It is clear
that such a physical situation can be found, as well, in the
macroworld. We have shown this based in our examples. We can now
ask why nonclassical probabilities only appeared in the
microworld. In the light of our hypothesis, the answer would be
that the type of measurement, introducing non-classical
probabilities, is never used to describe a macroscopic system,
because we have enough other measurements to replace them. This is
no longer the case in the microworld." D. Aerts (\cite{Aerts86},
p.203)
\end{quotation}}

I would like to emphasize it is misleading to think that the
hidden measurement approach opens the door for a realistic (even
deterministic) description of the quantum entity. It should be
clearly stated that determinism is a presupposition of the model
and should not be regarded as welcome back to a classical
description. For me, the most interesting teaching of the `vessels
of water' experiment \cite{Aerts82} is the fact that it presents a
classical system which has a non-kolmogorovian probability model.
The conclusion that quantum mechanics can have a deterministic
description should not be regarded as an embracing of the
Laplacean spirit, something which by no means can be justified,
not even in classical physics and which Cassirer \cite{Cassirer56}
has shown, leads to contradictions. The hidden measurements
approach has to do with our ignorance and the determination of the
model with which we want to approach reality, but it does not
discover the deterministic nor indeterministic character; rather,
it makes explicit the fact that the way in which we approach
nature determines our model of it\footnote{The hidden measurements
approach turns out to be, from the view of the complementary
descriptions approach, a {\it meta-description} which goes from a
classical description (determinism) to a quantum description
(indeterminism) by shifting a parameter (epsilon). It defines in
this way a set of continuous descriptions which lye in between the
quantum and the classical descriptions.}. Determinism and
indeterminism cannot be extended beyond the limits of experience
but must themselves be understood and defined as one of the
conditions of the possibility of experience; they do not follow in
any way from the description we choose to make about nature, but
are rather a consequence of it. What is really shown in the hidden
measurements approach is that the structure with which we choose
to describe a certain system determines in an explicit way what we
will find.

{\smallroman
\begin{quotation} ``We have found a strange footprint
on the shores of the unknown. We have devised profound theories,
one after another, to account for its origins. At last, we have
succeeded in reconstructing the creature that made the footprint.
And lo! It is our own." Sir Arthur Eddington \cite{Eddington20}
\end{quotation}}

\section*{Conclusions}

I have presented a view as a starting point to develop a
conception of science in which it is acknowledged the different
possibilities of a path with no end. The distinction between
different levels of descriptions evades the mistake of making
confusing relations between concepts of different incompatible
schemes. As we have shown in several examples the main creation of
paradoxes and weirdness of quantum mechanics comes from mixing
concepts which pertain to different levels of description.

I think it safe to say that the idea of understanding has been
deeply misunderstood, this misunderstanding has closed the doors
of the quantum theory. The complementary descriptions approach
presents an interpretation of quantum mechanics which turns its
attention to new concepts such as {\it ontological potentiality}.
I believe the development of such concepts will allow to present
more clearly the experiments which are being performed at present
with superposition states. We have been looking during several
decades through the key-hole of the atomic description, the
creation of new concepts is the missing key which can open the
doors of the quantum theory.

\section*{Acknowledgments}

I specially want to thank Diederik Aerts, Dennis Dieks and
Graciela Domenech for encouragement and guidance on my work. I am
indebt with Sven Aerts, Karin Verelst, Sonja Smets, Soazig Le
Bihan, Michiel Seevinck and Patricia Kauark for stimulating
discussions and comments on earlier drafts of this article.


\begin{thebibliography}{99}

\bibitem{Aerts81} Aerts, D. 1981, {\it The one and the many: towards a
unification of the quantum a classical description of one and many
physical entities}, Doctoral dissertation, Brussels Free
University, Brussels.

\bibitem{Aerts82} Aerts, D., 1982, ``Example of a macroscopical situation
that violates Bell inequalities", {\it Lett. Nuovo Cimento}, {\bf
34}, pp.107-111.

\bibitem{Aerts86} Aerts, D., 1986, ``A possible explanation for the
probabilities of quantum mechanics", {\it Journal of Mathematical
Physics}, {\bf 27}, pp.202-210.

\bibitem{Aerts88} Aerts, D., 1988, ``The entity and modern physics: the
creation-discovery view of reality" In {\it Interpreting Bodies:
Classical and Quantum Objects in Modern Physics}, E. Castellani
(ed.), Princeton University Press, Princeton.

\bibitem{AertsdeRondeBart} Aerts, D., de Ronde, C. and D'Hooghe B., 2005,
``Compatibility and Separability for Classical and Quantum
Entanglement", submitted to {\it International Journal of
Theoretical Physics}.

\bibitem{Agamben99} Agamben, G., 1999, {\it Potencialities}, Stanford University Press.

\bibitem{Bacciagaluppi96} Bacciagaluppi, G., 1996, {\it
Topics in the Modal Interpretation of Quantum Mechanics}, Doctoral
disserattion, University of Cambridge, Cambridge.

\bibitem{Bitbol00} Bitbol, M., 2000, ``Physique quantique et
cognition",
{\it Revue Internationale de Philosophie}, 2/2000, n. 212,
pp.299-328.

\bibitem{Bohm52a} Bohm, D., 1952, ``A suggested interpretations of
the quantum theory in therms of ``hidden variables": Part I", {\it
Physical Review 85}, pp.166-179.

\bibitem{Bohm52b} Bohm, D., 1952, ``A suggested interpretations of
the quantum theory in therms of ``hidden variables": Part II",
{\it Physical Review 85}, pp.180-193.

\bibitem{Bohr28} Bohr, N., 1928, ``The Quantum Postulate and the recent
development of atomic theory", {\it Nature}, {\bf 121},
pp.580-590.

\bibitem{Bohr34} Bohr, N., 1934, {\it Atomic Theory and the Description
of Nature}, Cambridge University Press, Cambridge.

\bibitem{Born71} Born, M., 1971, {\it The Born-Einstein Letters}, Walker and Company, New York.

\bibitem{Bub97} Bub, J., 1997, {\it Interpreting the Quantum World}, Cambridge
University Press, Cambridge.

\bibitem{Cassirer56} Cassirer, E., 1956, {\it Determinism and
Indeterminism in Modern Physics}, Yale University Press.

\bibitem{DallaChiaraGiuntini89} Dalla Chiara M.L. and Giuntini R.,
1989, ``Paraconsistent quantum logics", {\it Foundations of
Physics}, {\bf 19}, pp.891-904.

\bibitem{DallaChiaraGiuntini01} Dalla Chiara, M.L. and Giuntini, R., 2001,
``Quantum Logic", {\it Handbook of philosophical logic} (in
press); archive ref and link: quant-ph/0101028.

\bibitem{D'Espagnat76} D'Espagnat, B., 1976, {\it Conceptual Foundations
of Quantum Mechanics}, Benjamin, Reading MA.

\bibitem{D'Espagnat98} D'Espagnat, B., 1998, ``Quantum Theory: A
Pointer To An Independent Reality", archive ref and link:
quant-ph/9802046.

\bibitem{Dieks88a} Dieks, D., 1988, ``The Formalism of Quantum Theory: An
Objective description of reality", {\it Annalen der Physik}, {\bf
7}, Band 45, Heft 3, pp.174-190.

\bibitem{Dieks88b} Dieks, D., 1988, ``Quantum Mechanics and Realism",
{\it Conceptus XXII}, {\bf 57}, pp.31-47.

\bibitem{Dieks89} Dieks, D., 1989, ``Quantum Mechanics Without the
Projection Postulate and Its Realistic Interpretation", {\it
Foundations of Physics}, {\bf 19}, pp.1397-1423.

\bibitem{Dieks89b} Dieks, D., 1989,``Resolution of the Measurement
Problem through Decoherence of the Quantum State", {\it Physics
Letters A}, Vol 142, pp.439-446.

\bibitem{DieksPP} Dieks, D., ``Quantum Mechanics: an Intelligible
Description of Objective Reality?", Preprint.

\bibitem{DieksVermaas95} Dieks, D. and Vermaas, P., 1995, ``The Modal
Interpretation of Quantum Mechanics and Its Generalization to
Density Operators", {\it Foundations of Physics}, {\bf 25},
pp.145-158.

\bibitem{DieksVermaas98} Dieks, D. and Vermaas, P., 1998, {\it The Modal
Interpretation of Quantum Mechanics}, Dieks and Vermaas (eds.)
1998, Vol 60 of the Western Ontario Series in the Philosophy of
Science, Kluwer Academic Publishers, Dordrecht.

\bibitem{Dirac} Dirac, P.A.M., 1974, {\it The Principles of Quantum
Mechanics}, 4th Edition, Oxford University Press, London.

\bibitem{Eddington20} Eddington, A., 1920, {\it Space, Time, and
Gravitation}, Cambridge University Press.

\bibitem{Enz85} Enz, C.P., 1985, ``Wolfgang Pauli, Physicist and
Philosopher" In {\it Symposium on the Foundations of Modern
Physics 1985}, pp.241-255, P. Lahti and P. Mittelstaedt (eds.)
World Scientific, Johensuu.

\bibitem{Feynman92} Feynman, R.P., 1992, {\it The Character of Physical Law}, Penguin Books.

\bibitem{Finkelstein87} Finkelstein, D., 1987, ``All is flux" In {\it Quantum Implications: Essays in
honour of David Bohm}, pp.289-94, B.J. Hiley and F.D. Peat (eds.),
London: Routledge and Kegan Paul.

\bibitem{Folse87} Folse, H.J., 1987, ``Niels Bohr's Concept of
Reality" In {\it Symposium on the foundations of Modern Physics
1987}, pp.161-179, P. Lathi and P. Mittelslaedt (eds.), World
Scientific, Singapore.

\bibitem{Friedman02} Friedman, J., Patel, V., Chen, W., Tolpygo,
S. and Lukens, J., 2002, ``Quantum Superposition of Distinct
Macroscopic States", {\it Nature}, {\bf 406}, 43.

\bibitem{Heisenberg27} Heisenberg, W., 1927, ``Uber den
anschaulichen Inhalt der quantentheoretischen Kinematik und
Mechanic", {\it Zeitschrift fur Physik}, {\bf 43}, pp.172-98;
reprinted as ``The Physical Content of Quantum Kinematics and
Mechanics", translation by J.A. Wheeler and W.H. Zurek, in {\it
Quantum Theory and Measurement}, J.A. Wheeler and W.H. Zurek
(eds.).

\bibitem{Heisenberg49} Heisenberg, W., 1949, {\it The Physical Principles of the Quantum Theory}, Dover.

\bibitem{Heisenberg58} Heisenberg, W., 1958, {\it Physics and Philosophy},
World perspectives, George Allen and Unwin Ltd., London.

\bibitem{Heisenberg72} Heisenberg, W., 1972, {\it Dialogos sobre la F\'isica At\'omica},
Biblioteca de Autores Cristianos de la Editorial Cat\'olica,
Madrid.

\bibitem{Hendry84} Hendry, J., 1984, {\it The Creation of Quantum
Mechanics and the Bohr-Pauli Dialogue}, D. Reidel Publishing
Company, Dordrecht.

\bibitem{Howard89} Howard, D., 1989, ``Holism, Separability
and the Metaphysical implications of the Bell inequalities" In
{\it Philosophical Consequences of Quantum Theory: Reflections on
Bell's Theorem}, pp.224-253, Cushing and McMullin (eds.),
University of Notre Dame Press, Notre Dame, Indiana.

\bibitem{KochenSpecker67} Kochen, S. and Specker, E., 1967, ``The
Problem of Hidden Variables in Quantum Mechanics", {\it Journal of
Mathematics and Mechanics}, {\bf 17}, pp.59-87.

\bibitem{Laurikainen88} Laurikainen, K.V., 1988, {\it Beyond
the Atom, The Philosophical Thought of Wolfgang Pauli},
Springer-Verlag, Berlin, Heidelberg.

\bibitem{Pauli94} Pauli, W., 1994, {\it Writings on Physics and
Philosophy}, Enz, C. and von Meyenn, K. (eds.), Springer Verlag.

\bibitem{Peres93} Peres, A., 1993, {\it Quantum Theory: Concepts
and Methods}, Kluwer Academic Publishers, Dordrecht.

\bibitem{deRonde03} de Ronde, C., 2003, Master Thesis: {\it Perspectival
Interpretation of Quantum Mechanics (a story about correlations
and holism)}, Institute for History and Foundations of
Mathematical and the Natural Sciences, Utrecht University and
University of Buenos Aires, URL =
http://www.vub.ac.be/CLEA/people/deronde/.

\bibitem{deRonde04} de Ronde, C., ``Interpretaci\'on
perspectival de la mec\'anica cu\'antica y descripciones
complementarias" In {\it Volumen 10 de Epistemolog\'ia e Historia
de la Ciencia}, pp.161-167, Garcia, P. and Morey, P. (eds.),
Universidad Nacional de Cordoba, Cordoba, URL =
http://www.vub.ac.be/CLEA/people/deronde/.

\bibitem{deRonde05a} de Ronde, C., ``Potencialidad ontol\'ogica y teor\'ia cu\'antica" {\it
Volumen 11 de Epistemolog\'ia e Historia de la Ciencia},
Universidad Nacional de Cordoba, Cordoba, (submitted).

\bibitem{deRondeCDI}de Ronde, C., 2005, ``Complementary Descriptions (PART I):
A Set of Ideas Regarding the Interpretation of Quantum Mechanics",
archive ref and link: quant-ph/0507105.

\bibitem{deRonde05b} de Ronde, C., ``The Prespectival Interpretation Revisited", In preparation.

\bibitem{deRonde05c} de Ronde, C., ``Ontological Potentilaity in the Modal Interrpetation of Quantum Mechanics", In preparation.

\bibitem{Rovelli96} Rovelli, C., 1996, ``Relational Quantum Mechanics", archive ref and link: quant-ph/9609002.

\bibitem{Schrodinger35} Schr\"odinger, E., 1935, ``The
Present Situation in Quantum Mechanics", {\it Naturwiss}, {\bf
23}, p.807, translated to english in {\it Quantum Theory and
Measurement}, J.A. Wheeler and W.H. Zurek (ed.), Princeton Univ
Press 1983.

\bibitem{Michiel04} Seevinck, M., 2004, ``Holism, Physical Theories and Quantum
Mechanics", to be published in {\it Studies in History and
Philosophy of Modern Physics}, Preprint.

\bibitem{Teller85} Teller, P., 1985, ``Relational Holism and Quantum
Mechanics", {\it British Journal for the Philosophy of Science},
{\bf 37}, 71.

\bibitem{Teller89} Teller, P., 1989, ``Relativity, Relational Holism, and the
Bell inequalities" In{\it Philosophical Consequences of Quantum
Theory: Reflections on Bell's Theorem}, pp.208-223 in Cushing and
McMullin (eds.), University of Notre Dame Press, Notre Dame,
Indiana.

\bibitem{VanFraassen80} Van Fraassen, B.C., 1980, {\it The Scientific Image}, Clarendon Press, Oxford.

\bibitem{VanFraassen91} Van Fraassen, B.C., 1991, {\it Quantum
Mechanics: An Empiricist View}, Clarendon, Oxford.

\bibitem{Vermaas99} Vermaas, P.E., 1999, {\it A Philosophers Understanding
of Quantum Mechanics}, Cambridge University Press, Cambridge.

\bibitem{vWeizsacker51} Von Weizs\"acker, C.F., 1951, ``Kontinuit\"at
und M\"oglichkeit", {\it Die Natuurwissenschaften}, {\bf 38},
pp.533-543. Translated as ``Continuidad y Posibilidad" in {\it La
Imagen F\'isica del Mundo}, 1974, Biblioteca de Autores
Cristianos, Madrid.

\bibitem{vWeizsacker85} Von Weizs\"acker, C.F., 1985,
``Heisenberg's philosophy" In {\it Symposium on the Foundations of
Modern Physics 1985}, pp.277-293, P. Lathi and P. Mittelstaedt
(eds.), World Scientific, Singapore.

\bibitem{WheelerZurek83} Wheeler, J.A. and Zurek, W.H., 1983, {\it Quantum Theory and
Measurement}, J.A. Wheeler and W.H. Zurek (eds.), Princeton
University Press, New Jersey.

\bibitem{Wittgenstein} Wittgenstein, L., 1974, {\it Tractatus Logico
Philosophicus}, Routledge Classics, London.


\end{thebibliography}
\end{document}